\documentclass{PoS}
\usepackage{epsfig,subfigure,bm}
\def \beq{\begin{equation}}
\def \eeq{\end{equation}}
\def \beqa{\begin{eqnarray}}
\def \eeqa{\end{eqnarray}}
\def \lt{\left}
\def \rt{\right}
\def \c{\chi}
\def \cb{\bar\chi}
\def \bx{\bm x}
\def \btx{{\bm{\tilde x}}}
\def \cM{{\mathcal M}}
\def \unitmatrix{1\!\!\!1}

\def \eg{{\sl e.g.\ }}
\def \ie{{\sl i.e.\ }}
\def \viz{{\sl viz.\ }}
\def \etal{{\sl et al.\ }}
\def \ibid{{\sl ibid.\ }}

\def \np{\emph{Nucl.\ Phys.\ }}

\def \pl{\emph{Phys.\ Lett.\ }}
\def \pr{\emph{Phys.\ Rev.\ }}

\def \prl{\emph{Phys.\ Rev.\ Lett.\ }}

\def \pos{{PoS\ }}
\title{Screening masses of mesons in 2+1 flavour QCD\\ 
       \hfill {\tiny BI-TP 2007/25} \\ \vspace{-1cm} }
\ShortTitle{Screening masses of mesons in 2+1 flavour QCD}
\author{\speaker{Swagato Mukherjee} for RBC-Bielefeld Collaboration \\
        Fakult\"at f\"ur Physik,\\
        Universit\"at Bielefeld,\\
        D-33615 Bielefeld, Germany.\\
        E-mail: \email{smukher@physik.uni-bielefeld.de}  }
\abstract{
We present results for screening masses of light and strange mesons in $2+1$
flavour QCD using improved (\emph{p4fat3}) staggered fermions on $6\times24^3$
lattices. We have studied the screening masses of scalar, pseudo-scalar, vector and
axial-vector mesons along the line of constant physics, determined by a pion mass
$\approx220$ MeV and a kaon mass $\approx500$ MeV. In order to investigate the
cut-off and volume dependencies we have also performed studies of the meson
screening correlators in the non-interacting theory using the \emph{p4} and the
standard staggered discretizations. 
}
\FullConference{The XXV International Symposium on Lattice Field Theory\\
		July 30-4 August 2007\\
		Regensburg, Germany }
\begin{document}
\section{Introduction} \label{sc.intro}

To acquire a detailed knowledge about the nature of the Quark Gluon Plasma (QGP) it
is essential to study the in-medium properties of hadronic excitations. These studies
provide information about the important length-scales in the QGP. This in turn gives
an idea about the relevant degrees of freedom in the QGP and their possible physical
effects. Furthermore, these studies also illuminate crucial aspects of the chiral and
the $U_A(1)$ axial symmetry restorations in QCD. Since in the finite temperature
lattice QCD simulations the temporal extent of the lattice is limited by the inverse
of the temperature most of the finite temperature lattice studies have been
concentrated on spatial correlation function of hadrons. The exponential decay of
these spatial correlators defines the so called \emph{screening masses} \cite{detar}.
Physical interpretation of this quantity is as follows--- if one puts a test hadron
in the QGP medium then the spatial distance beyond which its effects are effectively
screened is given by the inverse of its screening mass. For some recent lattice
results on the screening masses of mesons see Ref.\ \cite{recent-lattice} and for a
review of the earlier results see Ref.\ \cite{karsch}. Here we present the first ever
lattice results for the meson screening masses in $2+1$ flavour QCD with realistic
quark masses.

\section{Operators} \label{sc.operators}

The lattice staggered meson operators are defined as 
$\cM(\bx) = \bar\psi(\bx)\lt(\Gamma^D\otimes\Gamma^F\rt)\psi(\bx)$,
where $\psi(\bx)$ is the fermion field at Euclidean space-time $\bx=(x,y,z,\tau)$.
The matrices $\Gamma^D$ and $\Gamma^F$ are products of $\gamma$-matrices and generate
the spin-flavour structure of the corresponding meson. Here we are interested in the
local meson operators, for which $\Gamma^D=\Gamma^F\equiv\Gamma$. For the staggered
fermions the local meson operators can be written as 
$\cM(\bx) = \tilde\phi(\bx) \cb(\bx)\c(\bx)$,
where $\tilde\phi(\bx)$ is a phase factor depending on the choice of $\Gamma$. 

We consider only the connected part of the screening correlators, \ie we
consider only the flavour non-singlet states. The connected part of the staggered
meson screening correlator, projected to zero momentum, can be obtained as
\beq
C(z) = \sum_\btx \phi(\bx)  
       \lt\langle\lt(M^{-1}_{{\bx}\bm{0}}\rt)^\dagger M^{-1}_{{\bm 0}\bx} \rt\rangle, 
\eeq
where $M^{-1}_{{\bm 0}\bx}$ is the full staggered propagator (\ie for $N_F=4$) from
${\bm 0}$ to $\bx$, $\phi(\bx)=(-1)^{x+y+\tau}\tilde\phi(\bx)$ and $\btx=(x,y,\tau)$.
Since a staggered fermion meson correlator, in general, contains two different mesons
with opposite parity \cite{golterman, altmeyer} we parametrize this correlator as
\beq
C(z) = A_{NO} \cosh\lt[ M^{scr}_- \lt( z - \frac{N_s}{2} \rt)\rt] +
       (-1)^z A_O \cosh\lt[ M^{scr}_+ \lt( z - \frac{N_s}{2} \rt)\rt] .
\label{eq.cor-par}       
\eeq
The parameters $M^{scr}_-$ and $M^{scr}_+$ are the screening masses
of the corresponding mesons. According to our convention $M^{scr}_-$
($M^{scr}_+$) corresponds to the screening masses of the lightest negative
(positive) parity states and $A_{NO}\ge0$, $A_O\le0$ \footnote{We have chosen our
convention such that the Goldstone pion comes as the non-oscillating part of the
screening correlator and with positive amplitude.}.

For the staggered fermion there are 8 possible local meson operators \cite{golterman,
altmeyer}. In this work we have studied all 8 of them.  The corresponding phase
factors $\phi(\bx)$ are listed them in Table\ \ref{tb.phase_factors}.

\begin{table}[!ht]
\begin{center}
\begin{tabular}{c c|c c|c c|c c} \hline
  Operators& $\phi(\bx)$& \multicolumn{2}{c|}{$\Gamma$}&
  \multicolumn{2}{c|}{$J^{PC}$}& \multicolumn{2}{c}{Physical states} \\
  && NO& O& NO& O& NO& O \\ \hline

  $\cM1$& $(-1)^{x+y+\tau}$& $\gamma_3\gamma_5$& $\unitmatrix$& $0^{-+}$& $0^{++}$&
  $\pi_2$& $a_0$ \\

  $\cM2$& $1$& $\gamma_5$& $\gamma_3$& $0^{-+}$& $0^{+-}$& $\pi^\pm$& -- \\

  $\cM3$& $(-1)^{y+\tau}$& $\gamma_1\gamma_3$& $\gamma_1\gamma_5$& $1^{--}$& $1^{++}$&
  $\rho^T_2$& $a_1^T$ \\ 

  $\cM4$& $(-1)^{x+\tau}$& $\gamma_2\gamma_3$& $\gamma_2\gamma_5$& $1^{--}$& $1^{++}$&
  $\rho^T_2$& $a_1^T$ \\

  $\cM5$& $(-1)^{x+y}$& $\gamma_4\gamma_3$& $\gamma_4\gamma_5$& $1^{--}$& $1^{++}$&
  $\rho^L_2$& $a_1^L$ \\

  $\cM6$& $(-1)^x$& $\gamma_1$& $\gamma_2\gamma_4$& $1^{--}$& $1^{+-}$& $\rho^T_1$&
  $b_1^T$ \\

  $\cM7$& $(-1)^y$& $\gamma_2$& $\gamma_1\gamma_4$& $1^{--}$& $1^{+-}$& $\rho^T_1$&
  $b_1^T$ \\

  $\cM8$& $(-1)^\tau$& $\gamma_4$& $\gamma_1\gamma_2$& $1^{--}$& $1^{+-}$& $\rho^L_1$&
  $b_1^L$ \\ \hline 
\end{tabular}
\end{center}
\caption{A complete list of the meson operators studied in this work (see \eg Table 1
of Ref.\ \cite{altmeyer}, with $z\leftrightarrow\tau$ interchanged). Possible
assignments of the corresponding physical states are given only for the $u$-$d$
flavours.}
\label{tb.phase_factors}
\end{table}

\section{Results} \label{sc.reuslts}

For the present study we have used the gauge configurations generated by the
RBC-Bielefeld collaboration using the \emph{p4fat3} staggered action on lattices of
size $N_\tau\times N_s^3=6\times24^3$. We have studied the meson screening
correlators along the Line of Constant Physics (LoCP) defined by the zero temperature
pion mass $m_\pi\approx220$ MeV and the zero temperature kaon mass $m_K\approx500$
MeV. (For more details about the simulation parameters and procedure see Ref.\
\cite{heide}). For each meson channel we have investigated the screening masses in
three different sectors corresponding to three different combination of quark masses,
\viz the $\bar ud$, $\bar us$ and $\bar ss$ sectors.

In a staggered fermion meson screening correlator a meson is always accompanied by an
opposite parity meson (see Eq.\ [\ref{eq.cor-par}]). In our present study, although
we have seen the signatures of both these mesons at our lowest temperature ($145$
MeV) the amplitude of one of them died out very fast with the increase of
temperature. Hence, in most of the cases we found the signature of only one of the
mesons. We summarize these findings below--- {\bf (I)} In the $\cM1$ channel we have
seen signature of only the positive parity Scalar (SC), and not the negative parity
non-Goldstone Pseudo-Scalar (PS). We will denote the screening masses of this SC
channel as $M^{scr}_{SC}$.  {\bf (II)} In $\cM2$ we have seen the signature of the
Goldstone PS, the only state present in this channel. We will denote the screening
masses of this channel by $M^{scr}_{PS}$.  {\bf (III)} In $\cM3$ and $\cM4$ we have
found signatures of only the positive parity Axial-Vector (AV). We have also found
that the screening masses of the AV (denoted by $M^{scr}_{AV}$ later) coming from
these two channels are degenerate for all temperatures.  {\bf (IV)} In $\cM5$ and
$\cM6$ we have only found the negative parity Vector (V) states. The screening masses
of these V states ($M^{scr}_V$) coming from these two channels are found to be
degenerate for all temperatures.  {\bf (V)} We found that $\cM5$ and $\cM8$ are very
noisy and reasonable signals were obtained only at our three highest temperatures. At
these temperatures we found signatures of both the V and AV states in these two
correlators. For these two channels, at our three highest temperatures (\viz $T=321$,
$363$ and $413$ MeV), we have found--- (a)
$M^{scr}_V(\cM5)=M^{scr}_{AV}(\cM8)=M^{scr}_{AV}=M^{scr}_V$, (b)
$M^{scr}_{AV}(\cM5)=M^{scr}_V(\cM8)>M^{scr}_{AV}=M^{scr}_V$. In fact, at high
temperatures all the V, AV states are not even expected to be degenerate. For the
spatial correlators at $T=0$ the rotational symmetry group of the lattice corresponds
to the continuum $O(3)$. At non-zero temperature this breaks down to $O(2)\times
Z(2)$. This, combined with the fact that $\cM5$ and $\cM8$ have longitudinal
polarizations compared to the transverse polarizations of the other V, AV channels,
explains the non-degeneracy of the V, AV states coming form $\cM5$ and $\cM8$
\cite{gupta}.

\begin{figure}[!t]
\begin{center}
\subfigure[]{\label{fig.t0_comp}\includegraphics[scale=0.55]{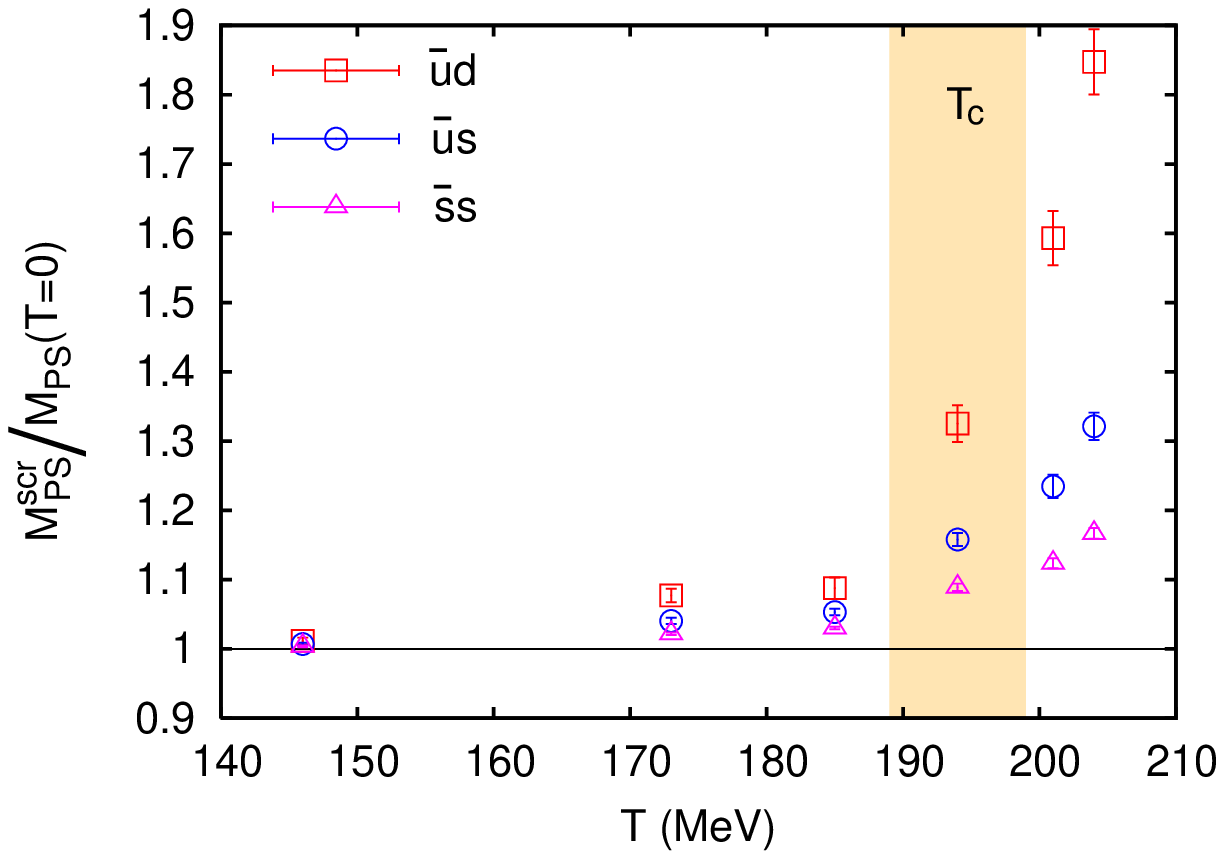}}
\subfigure[]{\label{fig.ps}\includegraphics[scale=0.55]{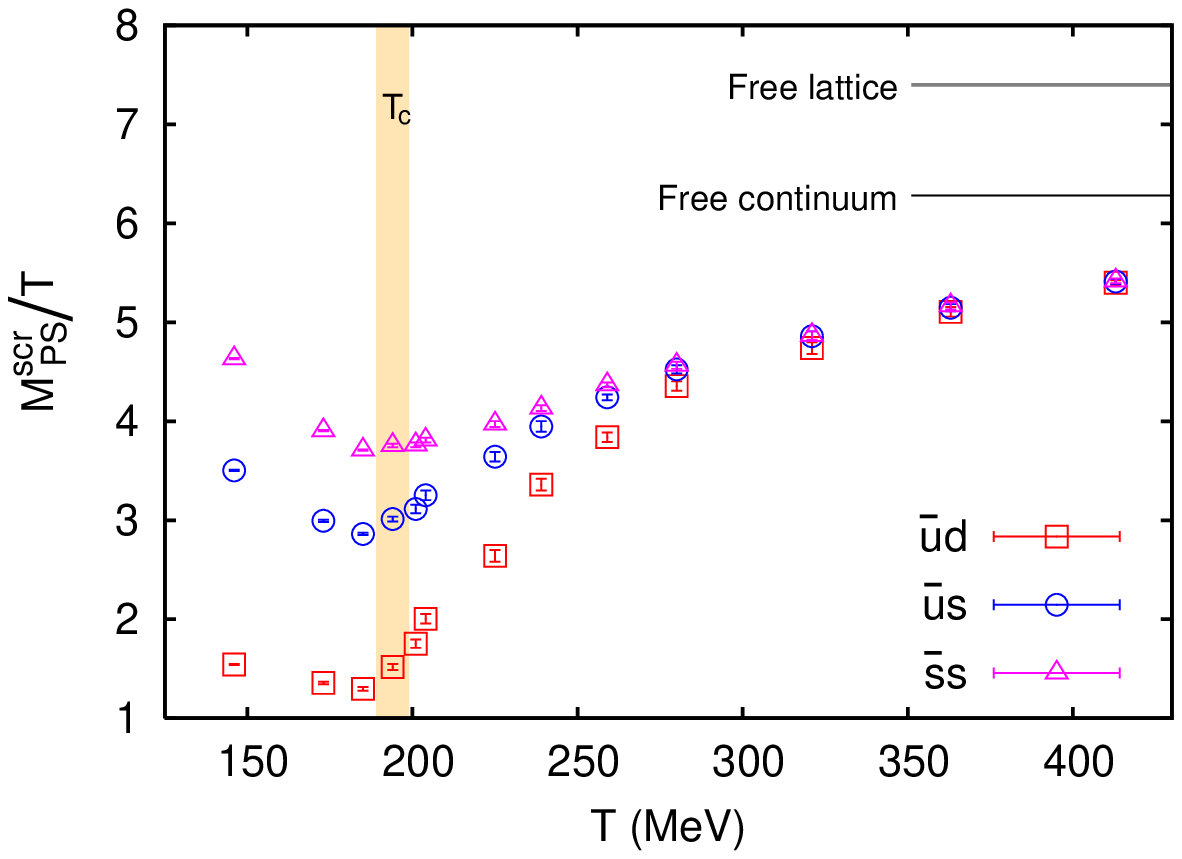}}
\subfigure[]{\label{fig.sc}\includegraphics[scale=0.55]{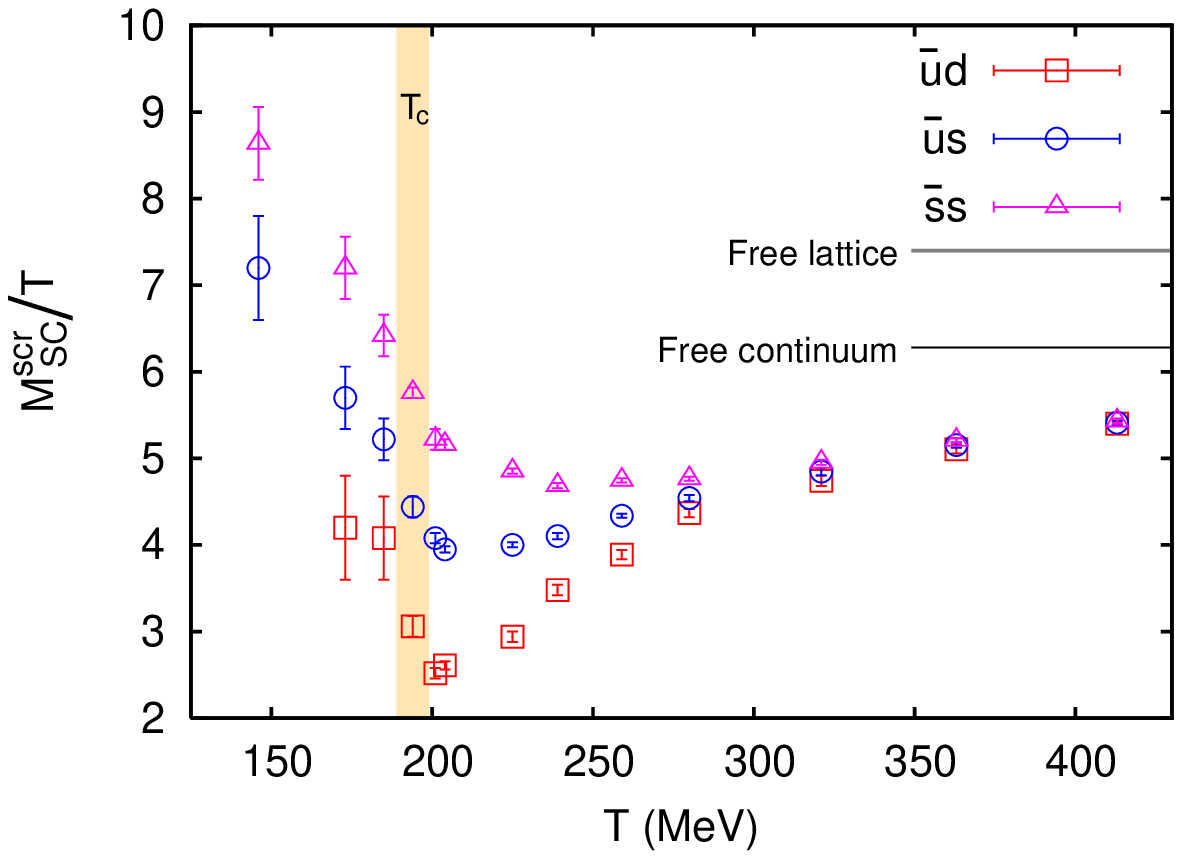}}
\subfigure[]{\label{fig.ps_sc}\includegraphics[scale=0.55]{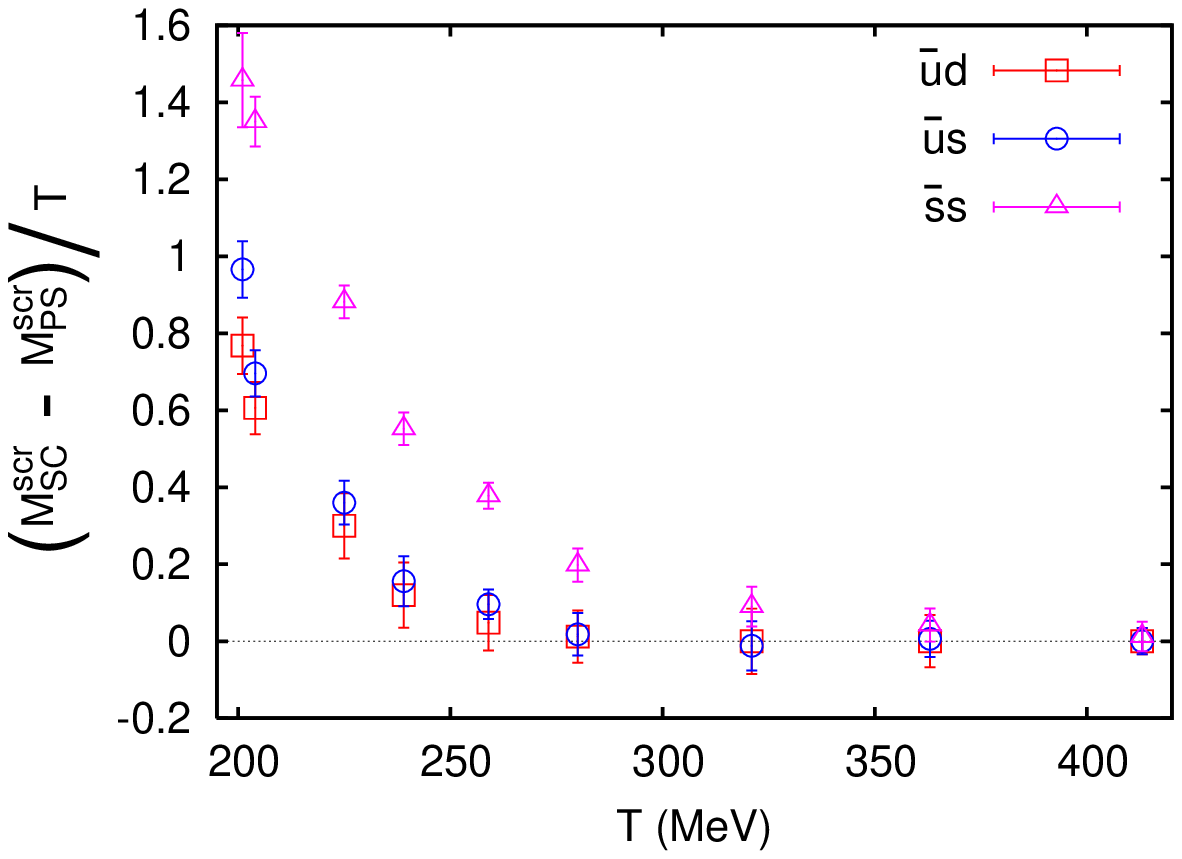}}
\end{center}
\vspace{-0.5cm}
\caption{(a) Deviation of the PS screening masses from the corresponding zero
temperature PS masses (determined at the same couplings along the LoCP). (b)
Temperature dependence of the PS screening masses. (c) Temperature dependence of the
SC screening masses. (d) Difference of the SC and PS screening masses as a function
of temperature. Shaded regions approximately indicate the transition region for the
\emph{p4fat3} staggered action on $6\times24^3$ lattice with $m_\pi\approx220$
MeV and $m_K\approx500$ MeV.}
\end{figure}

Since at zero temperature the screening masses are identical to the ordinary masses
it is interesting to investigate at what temperature their values start to differ. As
can be seen in Fig.\ \ref{fig.t0_comp} the ratio of $M^{scr}_{PS}$ to $M_{PS}(T=0)$,
the ordinary zero temperature PS masses determined at the same couplings along the
LoCP, starts differing from one for temperatures $T\gtrsim145$ MeV for all the three
quark sectors $\bar ud$, $\bar us$, $\bar ss$. In the range $145~{\rm MeV}\le T<T_c$
($\approx 194\pm5$ MeV), $M^{scr}_{PS}$ distinctly differs from $M_{PS}(T=0)$ but
only by $5-10\%$ depending on the quark mass. 

In Fig.\ \ref{fig.ps} and Fig.\ \ref{fig.sc} we show the temperature dependencies of
$M^{scr}_{PS}/T$ and $M^{scr}_{SC}/T$ respectively. Note that the minima of these two
screening masses (normalized by temperature) occur at different temperatures. This
probably indicates that the chiral symmetry (related to the minimum of
$M^{scr}_{PS}/T$) and the $U_A(1)$ axial symmetry (related to the minimum of
$M^{scr}_{SC}/T$) restorations are taking place at different temperatures. Even at
out largest temperature $T\gtrsim400$ MeV both $M^{scr}_{PS}$ and $M^{scr}_{SC}$
differ from their free continuum value of $2\pi T$ by $\sim10\%$ and even more
($\sim30\%$) from their ``free lattice values'' (see Section \ref{sc.free}). It is
expected, at least in the chiral limit, that $M^{scr}_{PS}$ and $M^{scr}_{SC}$ will
be degenerate once the $U_A(1)$ symmetry is effectively restored. In Fig.\
\ref{fig.ps_sc} we plot the difference between $M^{scr}_{SC}$ and $M^{scr}_{PS}$ as a
function of temperature. For the sectors containing the light quarks $M^{scr}_{PS}$
and $M^{scr}_{SC}$ become degenerate only at $T>250$ MeV, which is significantly
higher than the transition temperature $T_c$, and for the $\bar ss$ sector this
happens at an even higher temperature. These observations indicate that the effective
restoration of $U_A(1)$ probably does not take place at the transition temperature.

\begin{figure}[!t]
\begin{center}
\subfigure[]{\label{fig.v}\includegraphics[scale=0.55]{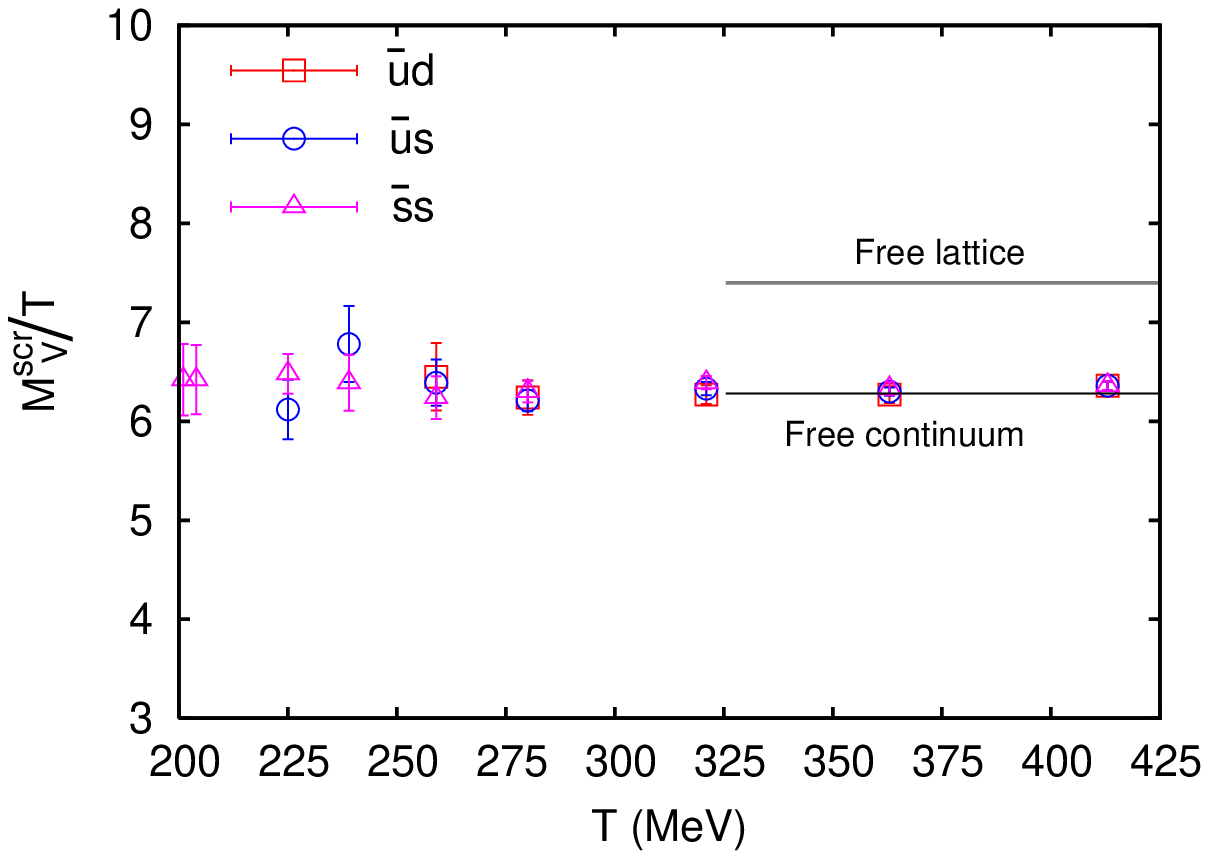}}
\subfigure[]{\label{fig.av}\includegraphics[scale=0.55]{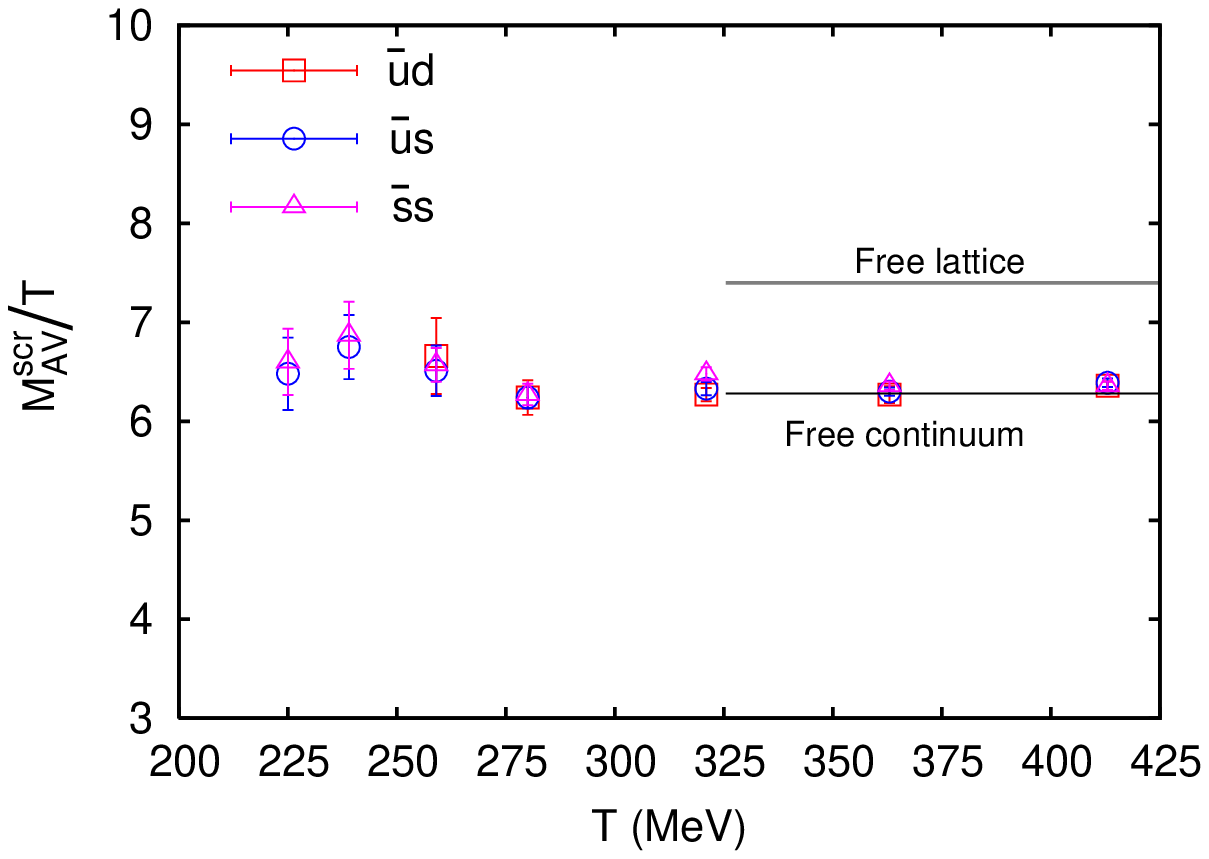}}
\subfigure[]{\label{fig.ps_v}\includegraphics[scale=0.55]{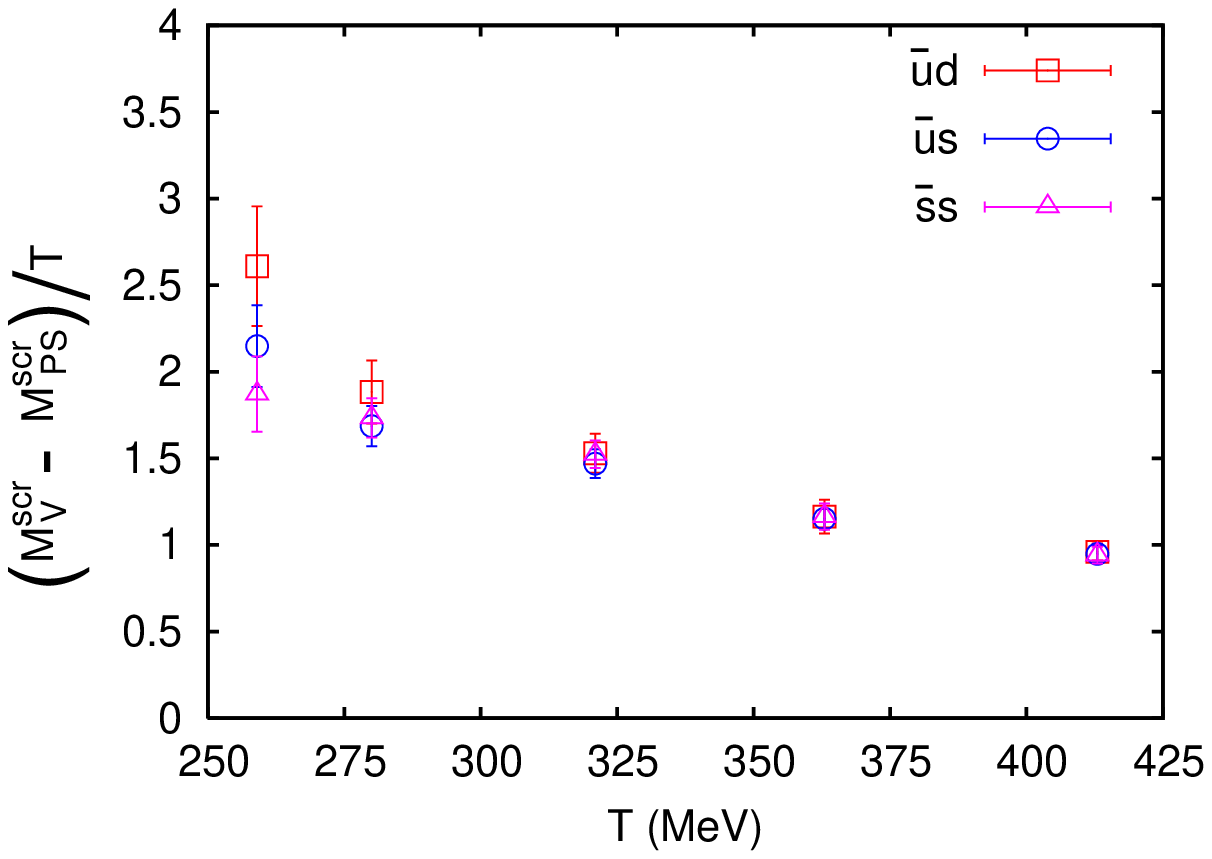}}
\subfigure[]{\label{fig.corr_comp}\includegraphics[scale=0.55]{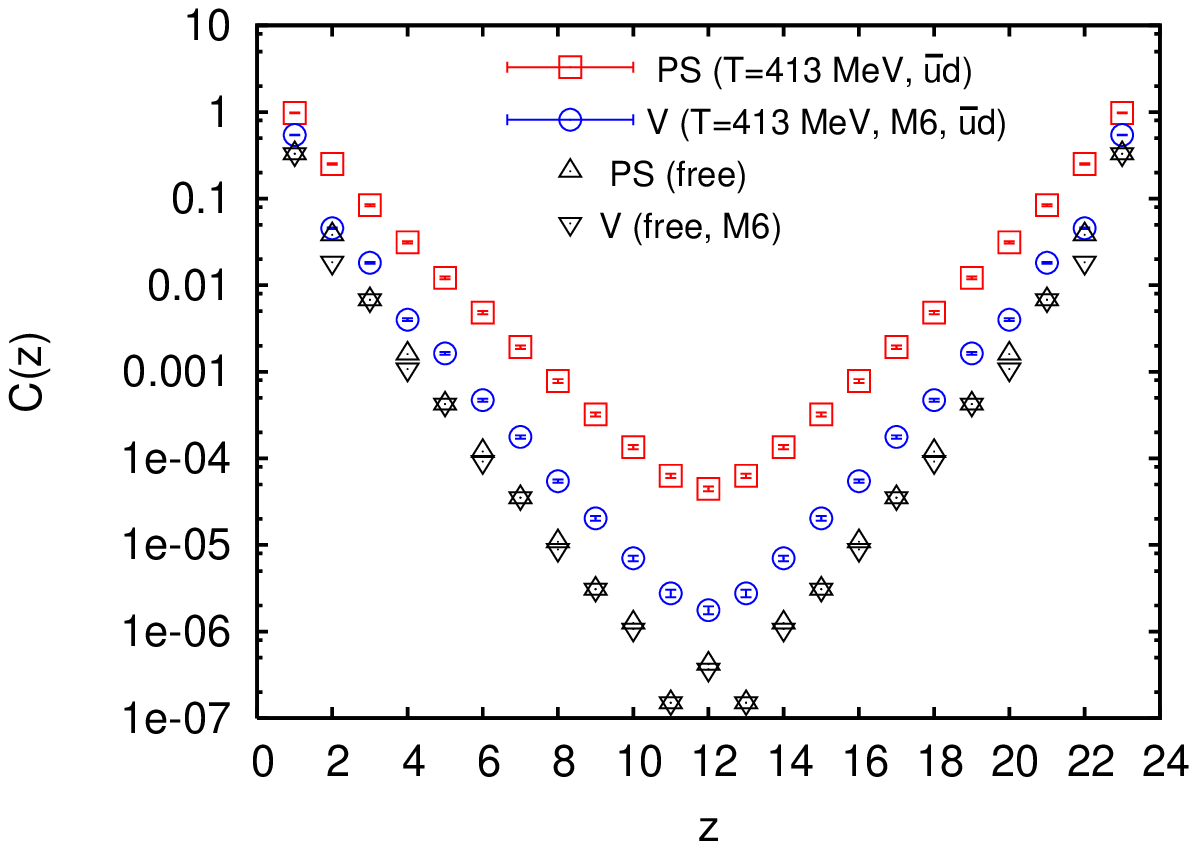}}
\end{center}
\vspace{-0.5cm}
\caption{(a) Temperature dependence of the screening masses of the V channel. (b)
Temperature dependence of the screening masses of the AV channel. (c) Difference
between the screening masses in the V and PS channel. (d) Comparison between the V
and PS meson screening correlators for the interacting case and also for the free
theory on an identical lattice.}
\end{figure}

In Fig.\ \ref{fig.v} and Fig.\ \ref{fig.av} we show the temperature dependencies of
$M^{scr}_{V}$ and $M^{scr}_{AV}$ respectively. We have found that above the
transition temperature both $M^{scr}_{V}$ and $M^{scr}_{AV}$ are, within errors,
compatible with their free continuum value of $2\pi T$ for all the three quark
sectors. Within our errors they are also degenerate with each other for the whole
temperature range $T>T_c$. However, they differ by $\sim20\%$ from their free lattice
values.

Via Fig.\ \ref{fig.ps_v} we investigate the degeneracy of $M^{scr}_{PS}$ and
$M^{scr}_{V}$, which is expected to take place at least in the limit of infinite
temperature. We have found that the difference between the $M^{scr}_{V}$ and
$M^{scr}_{PS}$ is far from zero at temperatures $T>400$ MeV ($>2T_c$). Moreover, our
analysis has shown that this difference is independent of the quark sectors for
$T>300$ MeV. In Fig.\ \ref{fig.corr_comp} we plot the screening correlators for the
PS and V channels of the $\bar ud$ sector and compare it with the corresponding
correlators for the free theory on an identical lattice. It is evident form this plot
that the V and PS correlators are degenerate for the free case, but for the
interacting theory it is clearly not so. This indicates that the non-degeneracy of
$M^{scr}_{V}$ and $M^{scr}_{PS}$ is probably not a lattice artifact and possibly
arises due to the presence interactions. However, one has to keep in mind that
compared to the free case the cut-off and finite volume effects could, in principle,
be very different for the interacting theory.

\section{Free case} \label{sc.free}

\begin{figure}[!t]
\begin{center}
\subfigure[]{\label{fig.free_naive}\includegraphics[scale=0.55]{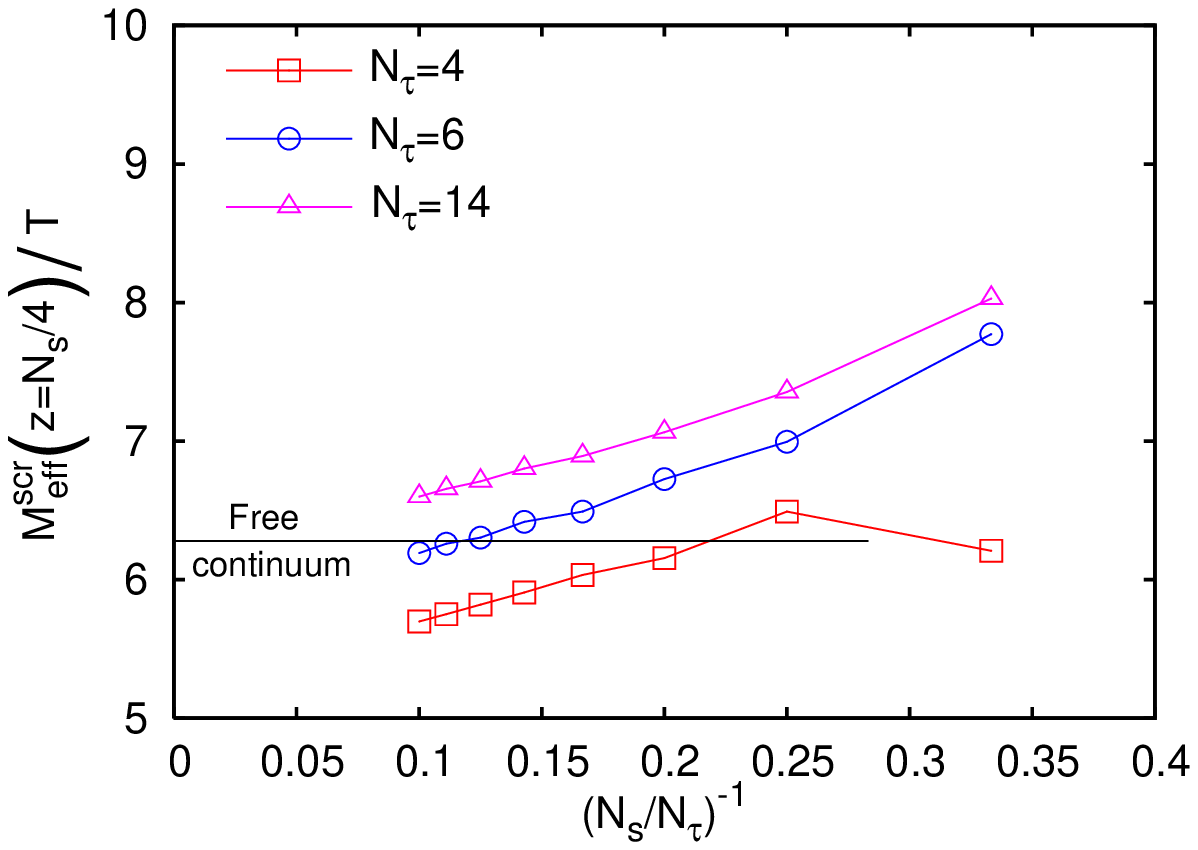}}
\subfigure[]{\label{fig.free_p4}\includegraphics[scale=0.55]{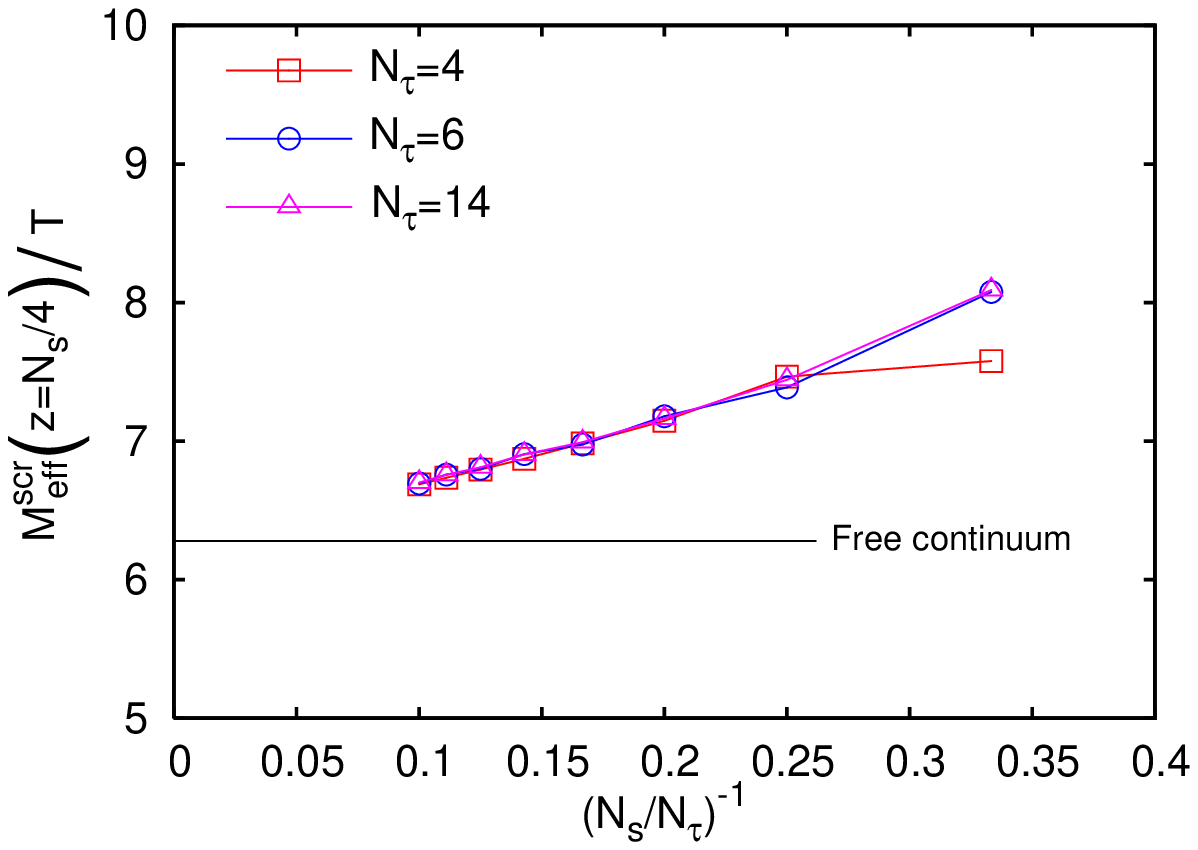}}
\end{center}
\vspace{-0.5cm}
\caption{(a) Dependencies of the effective screening mass (see Eq.\ [4.1]) of the PS
channel ($\cM2$) on $N_\tau$ and on the aspect ratio ($N_s/N_\tau$) for a free
(non-interacting) theory with standard staggered fermions.  (b) Same as (a) but with
\emph{p4} staggered action.}
\end{figure}

In order to have an idea about the cut-off and volume dependencies of our results we
have studied these meson screening correlators for the free (non-interacting) theory
using different lattice discretizations for the fermions, \viz the standard staggered
and the \emph{p4} fermions. Such studies can tell us about the lattice artifacts
already present in the free case. For the free case we computed the meson screening
correlators semi-analytically and looked at the effective mass of the PS ($\cM2$)
channel \footnote{In the free case we found that the screening masses of all the
channels are degenerate for all values of $N_\tau$ with $N_s/N_\tau\ge4$.}, defined
as---
\beq
aM^{scr}_{eff}(z) = - \ln\lt[ \frac{C(z+1)}{C(z)} \rt],
\label{eq.eff_mass}
\eeq
$a$ being the lattice spacing. Note that in the previous section the term ``free
lattice values'' means--- the value of $M^{scr}_{eff}(z=N_s/4)$ computed for the
non-interacting theory using \emph{p4} staggered fermions on a $6\times24^3$ lattice.

In Fig.\ \ref{fig.free_naive} and Fig.\ \ref{fig.free_p4} we show
$M^{scr}_{eff}(z=N_s/4)$ as a function of the inverse of the aspect ratio
($N_s/N_\tau$) for the standard staggered and the \emph{p4} fermions respectively. As
can be seen, for the improved \emph{p4} action there is almost no $N_\tau$ dependence
for $N_s/N_\tau\geq4$. On the other hand, for the unimproved standard staggered
action the discretization errors, \ie the $N_\tau$ dependence, are significantly
large.

However, for both the \emph{p4} and the standard staggered action we have found that
$M^{scr}_{eff}(z=N_s/4)$ is strongly dependent on the aspect ratio and the
corresponding continuum result of $2\pi T$ is reached only in the limit of infinite
volume. This suggests that the screening masses determined from lattice simulations
may have significant volume dependence. Hence a study of the volume dependence of the
screening masses is extremely important.

\section{Summary} \label{sc.summary}

We have investigated the screening masses of mesons in $2+1$ flavour QCD from $8$
different local meson operators (listed in Table\ \ref{tb.phase_factors}) and for
three different quark sectors $\bar ud$, $\bar us$ and $\bar ss$. For this purpose we
have used the gauge configurations, generated by the \emph{RBC-Bielefeld}
collaboration, using the improved \emph{p4fat3} fermion action on $6\times24^3$
lattices and along the LoCP determined by $m_\pi\approx220$ MeV and $m_K\approx500$
MeV.

We have found that in the PS channel the screening masses are identical to the
corresponding zero temperature (ordinary) masses for temperatures $T\lesssim145$ MeV.
In the high temperature regime both the PS and the SC screening masses differ from
the free continuum of $2\pi T$ by $\sim10\%$ even for $T>400$ MeV, which is larger
than twice the transition temperature. Moreover, for the $\bar ud$ and $\bar us$
sectors the PS and SC screening masses become degenerate only at $T\gtrsim250$ MeV
and at an even higher temperature for the $\bar ss$ sector.

For the V and AV channels we have found that the screening masses are compatible,
within our errors, with each other and with $2\pi T$ for the whole temperature range
of $T>T_c$. However, they are $\sim20\%$ below form our estimation of their free
lattice values. In contrast to the free lattice results the screening masses of the V
and PS channels do not become degenerate even for $T>400$ MeV. All these features are
true for all the three quark sectors.

We have also investigated the meson screening correlators for the non-interacting
theory. Whereas the cut-off dependence of the screening masses are almost negligible
for the improved \emph{p4} fermions, for the standard staggered fermion formulation
it is quite large. However, in both cases the screening masses show very strong
volume dependence. This makes the study of finite volume effects in meson screening
correlators extremely important. We hope to address this issue in future. 


\end{document}